\documentclass[fleqn,usenatbib]{mnras}

\usepackage{newtxtext,newtxmath}

\usepackage[T1]{fontenc}

\DeclareRobustCommand{\VAN}[3]{#2}
\let\VANthebibliography\thebibliography
\def\thebibliography{\DeclareRobustCommand{\VAN}[3]{##3}\VANthebibliography}

\usepackage{graphicx}	
\usepackage{amsmath}	
\usepackage{amsfonts}
\usepackage{bm,bbm}
\usepackage{xcolor}
\usepackage{caption}
\usepackage{comment}
\usepackage{hyperref}
\hypersetup{unicode=true}

\def\pd{{\rm d}}

\newcommand{\R}{\mathbbm{R}}

\title[Marked Correlation Function]{Revisiting Marked Galaxy Clustering from a Joint Point Process Perspective}

\author[T.\ T.\ Takeuchi]{
Tsutomu T.\ Takeuchi,$^{1,2}$\thanks{E-mail: tsutomu.takeuchi.ttt@gmail.com (TTT)}
\\
$^{1}$Division of Particle and Astrophysical Science, Nagoya University, Furo-cho, Chikusa-ku, Nagoya, 464–-8602, Japan\\
$^{2}$The Research Center for Statistical Machine Learning, The Institute of Statistical Mathematics, 10-3 Midori-cho, Tachikawa, Tokyo 190–-8562, Japan
}

\date{Accepted XXX. Received YYY; in original form ZZZ}

\pubyear{\the\year{}}

\begin{document}
\label{firstpage}
\pagerange{\pageref{firstpage}--\pageref{lastpage}}
\maketitle

\begin{abstract}
Marked correlation functions, in which galaxy properties such as luminosity or stellar mass are treated as marks, are widely used to test models of galaxy formation. 
In astronomy, however, these statistics are typically implemented as summary measures that do not preserve the joint structure of mark pairs conditioned on separation. 
In this work, we formulate galaxies as points $(x,m)$ on the product space $\mathbb{R}^3\times\mathcal{M}$, where $x$ denotes position and $m$ a mark, and introduce the joint pair correlation function $g(r;m_1,m_2)$ as the fundamental quantity describing mark-dependent clustering. 
We further define a diagnostic quantity $\Delta_{\mathrm{ind}}(r;m_1,m_2)$ that locally quantifies deviations from the independence hypothesis relative to spatial clustering alone, thereby providing a projection-free description of which mark pairs are over- or underrepresented at a given separation scale. 
Within this framework, commonly used diagnostics such as the inhomogeneous cross-$J$ function are naturally interpreted as summary statistics obtained through averaging over mark sets and geometric-event–based reductions of the joint structure. 
This perspective clarifies that previously discussed marked effects, including assembly bias, correspond to projections of an underlying joint dependence, and that observationally accessible information is the existence of non-factorizable joint structure itself. 
The present formulation provides both a fundamental quantity and practical diagnostics for its characterization.
\end{abstract}




\section{Introduction}\label{sec:introduction}

The two-point correlation function is the standard statistic used to characterize galaxy clustering, summarizing spatial inhomogeneity based solely on galaxy positions. 
At the same time, it has long been recognized that galaxy properties such as luminosity, stellar mass, color, star formation rate (SFR), and morphology depend on the surrounding environment. 
This has motivated the development of marked correlation statistics, in which correlation functions are weighted by such marks \citep{2000ApJ...545....6B,2005MNRAS.364..796S,2006MNRAS.369...68S}. 
For example, the well-known result that more luminous galaxies cluster more strongly has been interpreted as luminosity-dependent bias and has been tested in detail using large survey data \citep{2001MNRAS.328...64N,2005ApJ...630....1Z,2011ApJ...736...59Z}.

In this context, \citet{2000MNRAS.318L..45S} introduced the joint probability distribution of mark pairs conditioned on separation, $P(m_1,m_2\mid r)$, providing an early step toward describing mark-dependent clustering. 
Subsequent work, beginning with \citet{2005MNRAS.364..796S}, implemented marked correlation statistics as ratios of weighted second-order moments and applied them widely within frameworks such as the halo model. 
Like second-order summary statistics in spatial statistics, typified by Ripley’s $K$ function and its variants \citep[e.g.,][]{Ripley1981SpatialStatistics}, these measures robustly detect deviations from an uncorrelated (Poisson-like) baseline.

However, traditional marked statistics are designed as low-dimensional summaries and do not reconstruct the full joint structure of mark pairs conditioned on separation. 
This limitation is closely related to assembly bias. Simulations show that halo clustering at fixed mass can depend on secondary properties \citep{2005MNRAS.363L..66G,2007MNRAS.377L...5G}, and such dependencies can propagate into galaxy samples \citep{2013MNRAS.435.1313H,2014MNRAS.443.3044Z}. 
Because marked correlation functions explicitly weight clustering by internal properties, they are sensitive to conditional dependencies that remain hidden to standard two-point measures \citep{2009MNRAS.395.2381W}. 
Empirical studies further indicate that properties such as formation history, color, or luminosity exhibit systematic clustering trends beyond those explained by halo mass alone \citep{2002A&A...387..778G,2006MNRAS.369...68S,2009MNRAS.392.1080S,2009MNRAS.392.1467S}, and models sharing the same two-point correlation function can nevertheless be distinguished using marked statistics \citep{2009MNRAS.395.2381W}. 
Together, these findings suggest that clustering cannot be fully characterized by spatial configuration or mass alone.

From the standpoint adopted here, such phenomena are more naturally interpreted as consequences of projecting an underlying joint structure of marks and spatial separation. 
In this sense, assembly bias is best regarded not as a directly measurable physical quantity, but as a manifestation of projection. 
The key empirical fact motivating marked statistics is therefore the breakdown of separability between spatial configuration and internal properties.

To make this explicit, we treat galaxies as points $y=(x,m)$ in the product space $\mathbb{R}^3\times\mathcal{M}$ and adopt the joint pair correlation function $g((x_1,m_1),(x_2,m_2))$ as the fundamental quantity describing mark-dependent clustering. 
Here $\mathcal{M}$ denotes the mark space, encompassing continuous variables such as absolute magnitude, discrete labels such as morphology, and multivariate galaxy properties. 
We interpret conventional marked correlation statistics as projections of this joint structure, thereby making explicit what is measured and what information is lost under projection. 
Our goal is not to estimate assembly bias itself, but to characterize the joint structure that enables such phenomena.

We further introduce the mark-pair distribution conditioned on separation, $p(m_1,m_2\mid r)$, together with a diagnostic quantity that locally characterizes deviations from the independence hypothesis relative to spatial clustering alone,
\begin{align}
    \Delta_{\mathrm{ind}}(r;m_1,m_2)= \ln\frac{g(r;m_1,m_2)}{g_X(r)\,p(m_1)p(m_2)}.
\end{align}
{Here $g_X(r)\equiv1+\xi_X(r)$ denotes the usual pair correlation function of the position-only point process, i.e. the clustering signal obtained when marks are ignored. 
Accordingly, $\Delta_{\rm ind}=0$ corresponds to mark independence, while nonzero values indicate coupling between marks and clustering.}
This provides a projection-free framework for directly describing which mark pairs are over- or underrepresented at a given separation scale. 
{For continuous or multivariate marks, the joint structure may be compressed through basis-projected coefficients $C_{ab}(r)\equiv \mathbb{E}[\phi_a(m_1)\phi_b(m_2)\mid r]$,
where $\{\phi_a\}$ is a chosen family of basis functions on mark space (see Sec.~\ref{sec:discussion} for details).}
Unlike traditional marked estimators, $\Delta_{\mathrm{ind}}$ directly visualizes the underlying joint structure.

In spatial statistics, methods such as the inhomogeneous cross-$J$ function diagnose deviations from mark independence under minimal stationarity assumptions. 
These approaches reduce the joint position–mark structure to a one-dimensional function of separation $r$. 
In contrast, we adopt the joint pair correlation $g(r;m_1,m_2)$ itself as the fundamental quantity, allowing explicit identification of which mark combinations contribute to observed deviations.

The structure of this paper is as follows. 
Section~\ref{sec:point_process_on_direct_product_space} introduces the basic quantities for point processes on product spaces and defines the joint pair correlation and its reduced representations. 
Section~\ref{sec:traditional_mark_correlation_as_projection} shows that traditional marked correlation statistics can be understood as projections of the joint correlation and discusses reweighting together with diagnostics based on $\Delta_{\mathrm{ind}}$. 
Section~\ref{sec:discussion} discusses implementation and connections to spatial statistics, and Section~\ref{sec:summary_outlook} outlines possible extensions of this framework.
Appendix~\ref{app:palm} provides supplementary explanations of the point-process and measure-theoretic concepts used throughout this paper.

\section{Point Processes on the Product Space and the Definition of the Joint Two-Point Correlation}
\label{sec:point_process_on_direct_product_space}

In this section we introduce the minimal theoretical framework required to treat galaxy distributions as marked point processes.
The goal here is not to develop a mathematically rigorous theory of point processes, but rather to clarify, in a manner consistent with the correlation functions and cell-count statistics used in astronomy, what is defined and what is being conditioned upon.
A more general formulation of correlation functions for finite point processes, including higher-order statistics and conditioning structure, has been developed in \citet{takeuchi2026finite_sample_window}.

\subsection{Basic quantities of a marked point process}

We represent a galaxy as
\begin{align}
y=(x,m)\in\mathbb{R}^3\times\mathcal{M},
\end{align}
{where $x\in\R^3$ denotes \emph{the three-dimensional} spatial position of the galaxy}, and $m$ is an associated mark, such as luminosity (absolute magnitude), stellar mass, SFR, or morphology.
The space $\mathcal{M}$ denotes the mark space and may include continuous variables,
discrete labels, or multivariate quantities.

In this representation, the galaxy distribution is described as a set of points on the product space $\mathbb{R}^3\times\mathcal{M}$,
and is probabilistically treated as a marked point process.
In the language of point processes, two quantities play a fundamental role.

{The first-order intensity (product density) $\rho^{(1)}(x,m)$ is defined as the ensemble mean density such that 
\begin{align}
\rho^{(1)}(x,m)\,\pd^3 x\,\pd m
\end{align}
gives the expected number of points (galaxies) in an infinitesimal neighborhood of $(x,m)$.}
This corresponds to the average density when both position and mark are specified simultaneously.

{
Similarly, the second-order product density $\rho^{(2)}((x_1,m_1),(x_2,m_2))$ is defined through the ensemble expectation of pair counts such that
\begin{align}
    \rho^{(2)}((x_1,m_1),(x_2,m_2))\,\pd^3x_1\,\pd m_1\,\pd^3x_2\,\pd m_2
\end{align}
gives the expected number of ordered pairs simultaneously occurring
in the corresponding infinitesimal neighborhoods.
}
From an astronomical viewpoint, this quantity describes the frequency with which galaxy pairs possessing specified properties appear in specified configurations.

\subsection{Joint pair correlation: definition and intuition}

{
We first introduce the pair correlation function on the product space
as
\begin{align}
    g\left((x_1,m_1),(x_2,m_2)\right)
    \equiv
    \dfrac{\rho^{(2)}\left((x_1,m_1),(x_2,m_2)\right)}{\rho^{(1)}(x_1,m_1)\,\rho^{(1)}(x_2,m_2)}.
    \label{eq:joint_g}
\end{align}
Under statistical homogeneity and isotropy, this reduces to a function
of separation and marks, $g(r;m_1,m_2)$. 
In terms of the marginal mark distribution $p(m)$ and mean density $\lambda$, this can be written as
\begin{align}
g(r;m_1,m_2)
=
\dfrac{\rho^{(2)}(r;m_1,m_2)}{\lambda^2 p(m_1)p(m_2)}.
\label{eq:marginal_mark_distribution}
\end{align}
Here the normalization by $\lambda^2 p(m_1)p(m_2)$ ensures consistency with the marginal mark distribution and guarantees that marginalization over marks recovers the position-only correlation function.
We regard $g(r;m_1,m_2)$ as the fundamental joint structure from which conventional marked statistics are obtained as projections.
This quantity expresses how much more (or less) frequently a pair with marks $(m_1,m_2)$ occurs at separation $r$ relative to the completely uncorrelated (Poisson) case.}

The corresponding correlation function is defined as
\begin{align}
    \xi\left((x_1,m_1),(x_2,m_2)\right)
    =
    g\left((x_1,m_1),(x_2,m_2)\right)-1.
    \label{eq:joint_xi}
\end{align}
It is important to emphasize that both $g$ and $\xi$ directly describe the \emph{joint correlation between position and mark}.
That is, a situation in which the clustering strength depends on galaxy properties is naturally expressed through the $m$-dependence of $g$.

\subsection{Reduction under homogeneity and isotropy: representation by separation $r$}

In many galaxy surveys, statistical homogeneity and isotropy hold approximately on sufficiently large scales.
In such cases, positional dependence can be reduced to the pair separation
$r\equiv |x_1-x_2|$.

The joint pair correlation can then be written as
\begin{align}
g(r;m_1,m_2)
&\equiv
\dfrac{\rho^{(2)}(r;m_1,m_2)}
{\rho^{(1)}(m_1)\rho^{(1)}(m_2)}, \\
\xi(r;m_1,m_2)
&=
g(r;m_1,m_2)-1.
\label{eq:reduced_g}
\end{align}
This represents how much more frequently galaxy pairs at separation $r$ with marks $(m_1,m_2)$ occur relative to the uncorrelated case.
This $r$-based representation will be important for comparison with conventional marked correlation functions introduced later.

\section{Conventional Marked Statistics as Projections (Moments)}
\label{sec:traditional_mark_correlation_as_projection}

In this section we clarify which information contained in the joint correlation $g(r;m_1,m_2)$ is retained by the marked statistics widely used in astronomy, and which information is discarded.

\subsection{Mark distribution conditioned on separation $r$}

When a galaxy pair at separation $r$ is selected uniformly at random,
the conditional distribution of the mark pair $(m_1,m_2)$ is defined as
\begin{align}
p(m_1,m_2\mid r)
\equiv 
\dfrac{\rho^{(2)}(r;m_1,m_2)}{\rho_X^{(2)}(r)}.
\label{eq:cond_mark_pair}
\end{align}
Here,
\begin{align}
\rho_X^{(2)}(r)
=
\iint
\rho^{(2)}(r;m_1,m_2)\,\pd m_1\,\pd m_2
\end{align}
is the second-order product density for positions only, obtained by marginalizing over marks.

The quantity $p(m_1,m_2\mid r)$ most directly expresses
how galaxy properties are distributed under the environmental condition defined by separation $r$,
and in our viewpoint constitutes the primary object of marked clustering.

\subsection{Weighted marked statistics as moments}

Introducing a commonly used weight $w(m)$,
we obtain
\begin{align}
\mathbb{E}\!\left[w(m_1)w(m_2)\mid r\right]
=
\iint
w(m_1)w(m_2)\,p(m_1,m_2\mid r)\,\pd m_1\,\pd m_2.
\label{eq:mark_moment}
\end{align}
This is a second moment of the conditional distribution $p(m_1,m_2\mid r)$.
Thus, conventional marked correlation functions can be understood as moment-based summaries of this conditional distribution.

Accordingly, conventional statistics retain only a \emph{limited portion}
of the information contained in $p(m_1,m_2\mid r)$ or equivalently in $g(r;m_1,m_2)$.

\subsection{Marked statistics as reweighted moments}

As discussed above, the joint pair correlation $g(r;m_1,m_2)$ is the fundamental quantity that fully encodes the relationship between marks and clustering, and conventional marked correlation functions can be viewed as its projections.
Here we reinterpret this projection operation from the viewpoint of \emph{reweighting} in order to clarify the relationship between conventional statistics and the framework developed in this work.

Introduce a nonnegative weight function $u(m)$ depending on the mark, and define the reweighted product densities on the product space as
\begin{align}
\rho_u^{(1)}(x,m)
&\equiv
u(m)\,\rho^{(1)}(x,m), \\
\rho_u^{(2)}((x_1,m_1),(x_2,m_2))
&\equiv
u(m_1)u(m_2)\,
\rho^{(2)}((x_1,m_1),(x_2,m_2)).
\label{eq:reweighted_densities}
\end{align}
This corresponds to modifying which marks (or pairs of marks) are treated as statistically representative.
{
This reweighting does not alter the underlying spatial configuration
of galaxies. 
Rather, it changes which galaxies (or galaxy pairs) are
statistically emphasized when forming projected summaries.
For example, luminosity weighting highlights more luminous galaxies without modifying the spatial distribution itself.}

Under this reweighting, the joint pair correlation becomes
\begin{align}
g_u(r;m_1,m_2)
=
\dfrac{\rho_u^{(2)}(r;m_1,m_2)}
{\rho_u^{(1)}(m_1)\rho_u^{(1)}(m_2)}
=
g(r;m_1,m_2),
\label{eq:invariance_of_g}
\end{align}
showing that \emph{the joint correlation itself is invariant under reweighting}.
On the other hand, conditional expectations at separation $r$,
\begin{align}
\mathbb{E}_u\!\left[1\mid r\right]
=
\dfrac{\displaystyle \iint u(m_1)u(m_2)\,\rho^{(2)}(r;m_1,m_2)\,\pd m_1\,\pd m_2}
{\displaystyle \iint \rho^{(2)}(r;m_1,m_2)\,\pd m_1\,\pd m_2},
\label{eq:reweighted_moment}
\end{align}
depend on the choice of $u(m)$.
Conventional marked correlation functions correspond to specific choices of such reweighted moments,
and can therefore be understood as statistics that retain only low-dimensional information about the joint structure.

\subsection{Direct measurement of deviation from the independence (separability) hypothesis}

The joint pair correlation $g(r;m_1,m_2)$ introduced above is the fundamental quantity that fully captures the joint structure between marks and clustering.
To interpret this information, however, it is necessary to specify the reference against which excess or deficit is evaluated.
Here we introduce a simple and intuitive diagnostic based on clustering in position alone.

Let the pair correlation for the position-only point process be written as
\begin{align}
g_X(r)\equiv 1+\xi_X(r).
\label{eq:gx_def}
\end{align}
This quantity represents how much more frequently galaxy pairs at separation $r$ occur relative to a Poisson distribution when marks are ignored.

Using this $g_X(r)$ as a reference, we define
\begin{align}
\Delta_X(r;m_1,m_2)
\equiv
\ln \dfrac{g(r;m_1,m_2)}{g_X(r)}.
\label{eq:Delta}
\end{align}
{This ratio is closely related in form to the conventional marked correlation function. However, it differs in that it is defined locally on the $(m_1,m_2)$ space without projection. }
{
Here, the conventionally defined marked correlation function is typically written as
\begin{align}
    M(r)=\frac{1+W(r)}{1+\xi_X(r)}, \label{eq:mark_correlation_classical}
\end{align}
where $W(r)$ represents the weighted correlation and $\xi_X(r)$ is the position-only correlation function.
}
{This conventional marked correlation functions are obtained only after integrating this joint structure over marks with a specified weight.}
This measures, on a logarithmic scale, how much more ($\Delta_X>0$) or less ($\Delta_X<0$) frequently galaxy pairs with marks $(m_1,m_2)$ occur at separation $r$ compared to the expectation from clustering in position alone.

More generally, to describe deviations from an arbitrary reference structure $g_0(r;m_1,m_2)$, we define
\begin{align}
\Delta_0(r;m_1,m_2)
\equiv
\ln \dfrac{g(r;m_1,m_2)}{g_0(r;m_1,m_2)}.
\end{align}

Under the mark-independence hypothesis
\begin{align}
p(m_1,m_2\mid r)=p(m_1)p(m_2),
\end{align}
the natural reference becomes
\begin{align}
    g_0(r;m_1,m_2)=g_X(r)p(m_1)p(m_2). \label{eq:g_baseline}
\end{align}
{This reference represents the mark-independent baseline, in which marks are assigned independently of position. 
In practice, it is equivalent to the correlation obtained by randomly shuffling marks among galaxies while preserving their spatial configuration.
Equation~\eqref{eq:g_baseline} yields}
\begin{align}
    \Delta_{\mathrm{ind}}(r;m_1,m_2)
    \equiv
    \ln \dfrac{g(r;m_1,m_2)}{g_X(r)p(m_1)p(m_2)}.
\end{align}

As a linearized measure of deviation in the weak-correlation regime,
we may also introduce the relative deviation from the reference structure:
\begin{align}
\mathfrak{D}_0(r;m_1,m_2)
\equiv
\dfrac{g(r;m_1,m_2)-g_0(r;m_1,m_2)}{g_0(r;m_1,m_2)}.
\label{eq:delta0_def}
\end{align}
The relation to the logarithmic diagnostic is then
\begin{align}
\Delta_0(r;m_1,m_2)
=
\ln\!\left(1+\mathfrak{D}_0(r;m_1,m_2)\right).
\label{eq:Delta_delta0_relation}
\end{align}
In the weak-correlation limit $|\mathfrak{D}_0|\ll1$,
\begin{align}
\Delta_0(r;m_1,m_2)\simeq \mathfrak{D}_0(r;m_1,m_2)
\end{align}
holds.
Thus $\mathfrak{D}_0$ may be regarded as a linearized deviation,
while $\Delta_0$ provides its logarithmic (nonlinear) representation.

If marks and clustering are unrelated, i.e., position and mark are strongly separable,
then
\begin{align}
g(r;m_1,m_2)\simeq g_X(r),
\label{eq:separation_null}
\end{align}
and hence for all $(m_1,m_2)$,
\begin{align}
\Delta_X(r;m_1,m_2)\simeq 0.
\label{eq:Delta_null}
\end{align}
Therefore, nonzero values of $\Delta_X$ directly indicate coupling between marks and spatial configuration, i.e., a breakdown of the separability hypothesis.

Importantly, $\Delta_X$ is not an integrated summary over mark space,
as in weighted moments,
but rather a localized diagnostic in $(m_1,m_2)$.
This allows direct visualization of
which combinations of marks are particularly overrepresented (or underrepresented)
at which distance scales.

\begin{figure}
    \centering
    \includegraphics[width=0.95\linewidth]{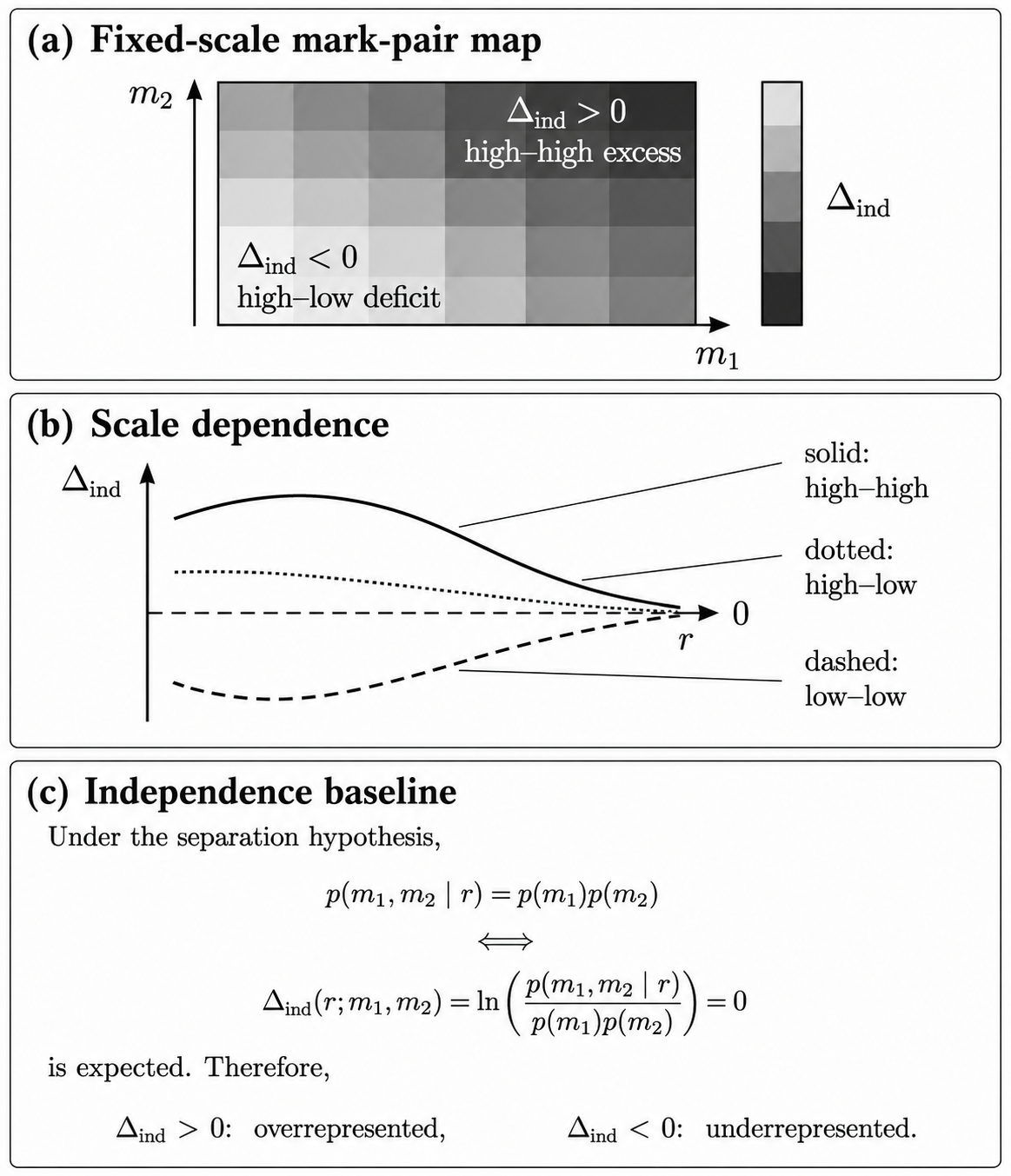}
    \caption{
    {Visualization of $\Delta_{\rm ind}(r;m_1,m_2)$ (schematic). 
    (a) A heatmap of $\Delta_{\rm ind}$ on the $(m_1,m_2)$ plane at a fixed $r=r_\ast$ directly shows which mark pairs are relatively enhanced or suppressed. 
    (b) Curves of $\Delta_{\rm ind}(r)$ for representative mark pairs summarize the scale dependence. 
    (c) $\Delta_{\mathrm{ind}}(r;m_1,m_2)\simeq 0$ corresponds to the separation (independence) hypothesis; nonzero $\Delta_{\rm ind}$ indicates a breakdown of factorization in the conditional mark-pair distribution.}
    }\label{fig:Delta_schematic}
\end{figure}

In this sense, $\Delta_X$ complements conventional marked correlation functions
and provides an effective tool for examining the detailed structure of mark-dependent clustering.
A schematic example of the visualization of $\Delta_X$ is shown in Fig.~\ref{fig:Delta_schematic}.

\subsection{Physical intuition: Reweighting as changing which galaxies are emphasized}

The reweighting operation introduced above may at first appear to be an abstract mathematical procedure.
From the standpoint of galaxy physics, however, it has a simple interpretation.
It changes which galaxies (or galaxy pairs) are regarded as representative when measuring clustering.

{
For example, let the mark represent galaxy luminosity $L$.
An analysis using the weight
\begin{align}
    u(L)=1
\end{align}
corresponds to the standard correlation function in which all galaxies are counted equally.
Using a weight proportional to luminosity,
\begin{align}
u \propto L 
\end{align}
corresponds to emphasizing more luminous galaxies.
}
This incorporates the question ``Do luminous galaxies cluster more strongly?'' directly into the definition of the correlation function.

Importantly, such reweighting does not change the spatial structure of the galaxy distribution itself.
The joint pair correlation $g(r;m_1,m_2)$ remains invariant.
What changes is which galaxy pairs are statistically emphasized in the projection.

\begin{figure}
    \centering
    \includegraphics[width=0.95\linewidth]{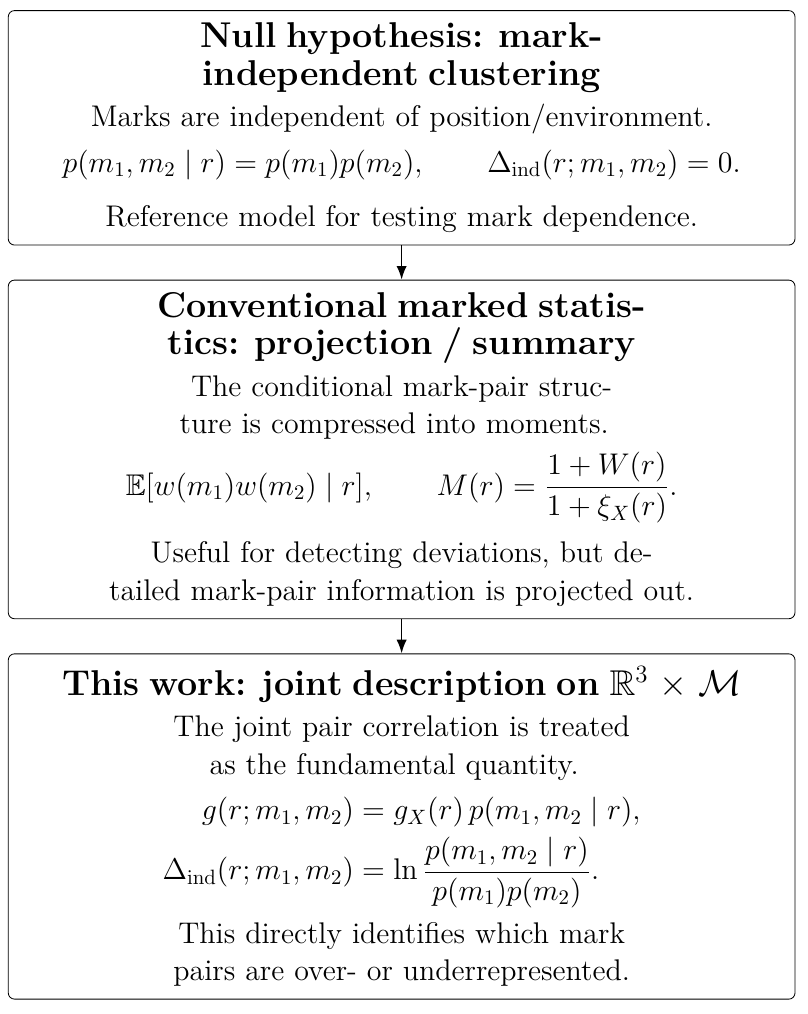}
    \caption{{
    Schematic comparison of three perspectives on marked clustering.
    Top: mark-independent clustering as the null hypothesis.
    Middle: conventional marked statistics as projected summaries.
    Bottom: the present work treats the joint pair correlation $g(r;m1,m2)$ as the fundamental quantity and diagnoses departures from independence through $\Delta_{\rm ind}(r; m1,m2)$.}
    }\label{fig:concept_null_projection_joint}
\end{figure}

From this viewpoint, conventional marked correlation functions may be understood as observing particular projections of the same underlying joint structure through weighting by physical properties such as luminosity, mass, or SFR.
Seemingly distinct effects, including size bias, luminosity bias, and SFR bias, are therefore unified as manifestations of the same joint correlation structure under projection.
{For clarity, we summarize the relation among the null hypothesis, conventional projected marked statistics, and the joint-structure perspective introduced in this work in Fig.~\ref{fig:concept_null_projection_joint}.}

In the language of point process theory, this reweighting corresponds to mathematically well-defined operations such as Palm measures and tilting (see Appendix~A).
From the standpoint of galaxy physics, however, it may be more intuitively understood as selecting which galaxies are emphasized when forming projected summaries of the joint mark--clustering structure.
In the following sections, this interpretation will be illustrated through concrete examples.

\section{Structural Interpretation in the Cosmological Context}

\subsection{Phenomenological motivation}

A common empirical finding in studies of galaxy and halo clustering is that internal properties are not randomly distributed with respect to spatial configuration.
Instead, quantities such as formation history, color, or luminosity exhibit systematic trends with environment.
This indicates that clustering cannot be fully characterized by spatial configuration or {halo mass alone.}
Importantly, such trends do not by themselves imply the presence of additional physical interactions between galaxy properties and environment.
Rather, they point to a breakdown of separability between spatial clustering and internal properties.
From an observational standpoint, the essential feature is therefore the existence of dependencies that cannot be captured by position-only statistics.

Marked statistics have traditionally been used to probe these effects by introducing weights based on internal properties.
However, the phenomenological role of such statistics is not to isolate specific physical mechanisms, but to reveal departures from the hypothesis that marks and spatial clustering are independent.
In this sense, the motivation for introducing marks is fundamentally structural:
they provide a means to test whether the joint distribution of position and internal properties factorizes.
The observational problem is thus not the identification of particular secondary physics, but the characterization of non-separable structure in galaxy populations.

\subsection{Structural origin of the problem}

Rather, they suggest that observed clustering statistics are projections of the joint structure involving internal properties.
Introducing a variable $m$ representing internal properties, a complete description of spatial dependence is given by $g(r;m_1,m_2)$.
The usual two-point correlation function can then be written as its marginalization:{
\begin{align}
    g_X(r) = \iint g(r;m_1,m_2)\, p(m_1)p(m_2)\,\pd m_1 \pd m_2.
\end{align}
Here the weighting by $p(m_1)p(m_2)$ arises from the normalization
used in the definition of $g(r;m_1,m_2)$ with respect to the marginal
mark distributions (eq.~\eqref{eq:marginal_mark_distribution}).
}

From this viewpoint, \citet{2002A&A...387..778G} showed that halo formation history affects clustering, \citet{2006MNRAS.369...68S} showed that mark-dependent segregation can be explained by halo occupation, \citet{2009MNRAS.392.1080S} attributed color correlations to central–satellite structure, and \citet{2009MNRAS.395.2381W} demonstrated degeneracy among models sharing the same $g(r)$.
All of these are consistent with the existence of non-factorization, i.e. a breakdown of independence between marks and spatial clustering:
{
\begin{align}
    g(r;m_1,m_2)\neq g_X(r)\,p(m_1)p(m_2).
\end{align}
Assembly bias can be expressed in this framework as
\begin{align}
    \Pr(\text{clustering} \mid M,\theta) \neq \Pr(\text{clustering} \mid M)
\end{align}
where $\Pr$ denotes the probability, and 
{{$\theta$} denotes a latent variable representing additional (unobserved) properties beyond the primary mark.}
Thus, the essential issue is not identifying specific physical dependencies, but recognizing that observed statistics are projections of a higher-dimensional joint structure.
}

\subsection{Projection structure for implementation and non-identifiability of assembly bias estimation}

We now summarize the discussion in a form directly connected to implementation.
The fundamental quantity is the bivariate mark structure conditioned on distance $r$, namely $g(r;m_1,m_2)$.
Conventional marked correlation functions can be uniformly expressed as projections of this quantity.

For example, specifying a pair weight $w(m_1,m_2)$ leads to the projected statistic
\begin{align}
    S_w(r)&=\dfrac{\displaystyle \iint w(m_1,m_2)\,\rho^{(2)}(r;m_1,m_2)\,\pd m_1\pd m_2}{\displaystyle \iint \rho^{(2)}(r;m_1,m_2)\,\pd m_1\pd m_2} \notag \\
    &= \iint w(m_1,m_2)\,p(m_1,m_2\mid r)\,\pd m_1\pd m_2.
\end{align}
Here $S_w(r)$ represents the expectation of the weight over pairs at distance $r$.
Ratio-type marked correlation functions commonly used in practice can be recovered as special cases through appropriate choices of $w$.
From this perspective, the question of which statistic to use becomes a design problem:
which weight $w$ should be chosen and which projection should be taken.

Diagnostics are defined relative to a reference model $g_0(r;m_1,m_2)$ via
\begin{align}
    \Delta_0(r;m_1,m_2)\equiv \ln\frac{g(r;m_1,m_2)}{g_0(r;m_1,m_2)}.
\end{align}
Deviations of any projected statistic can then be written using the conditional distribution under the reference model:
\begin{align}
    S_w(r) &- S_{w,0}(r) \notag \\
    & = \iint w(m_1,m_2)\,\left[p(m_1,m_2\mid r)-p_0(m_1,m_2\mid r)\right]\pd m_1\pd m_2.
\end{align}
In particular, under the logarithmic ratio $\Delta_0$,
\begin{align}
p(m_1,m_2\mid r)
=
\frac{p_0(m_1,m_2\mid r)\,\exp\!\left(\Delta_0(r;m_1,m_2)\right)}
{\displaystyle \iint p_0(m'_1,m'_2\mid r)\,\exp\!\left(\Delta_0(r;m'_1,m'_2)\right)\pd m'_1\pd m'_2}.
\end{align}
Thus $\Delta_0$ locally describes which regions of $(m_1,m_2)$ are relatively enhanced compared to the reference model.
This expression enables visualization of which mark pairs contribute to the signal of a given marked statistic,
and allows implementation to separate the design of $w$ from the estimation of $\Delta_0$.

Turning to assembly bias, it is often described as clustering depending on secondary variables {$\theta$} even at fixed mass $M$.
However, observationally we only access projected statistics resulting from latent generative processes involving {$\theta$}.
Introducing a latent-variable model,
\begin{align}
    g(r;m_1,m_2)=\iint g(r;m_1,m_2\mid {\theta_1},{\theta_2})\,p({\theta_1},{\theta_2})\,\pd {\theta_1}\pd {\theta_2},
\end{align}
the observable quantity is the marginalized $g(r;m_1,m_2)$,
and recovering individual {$\theta$}-dependence is generally impossible.

Moreover, the existence of counterexamples—different latent models yielding identical $g(r;m_1,m_2)$—
implies that observational statistics alone cannot uniquely identify {$\theta$}-dependence.
Therefore, \emph{estimators for ``assembly bias itself'' cannot be constructed without model assumptions.}
From our perspective, this non-identifiability arises not from missing physics,
but from the fact that observational statistics are projections of joint structure.
What can be robustly inferred is only the existence of non-factorization:
\begin{align}
    g(r;m_1,m_2)\neq g(r)\,p(m_1)p(m_2),
\end{align}
or equivalently $\Delta_{\mathrm{ind}}(r;m_1,m_2)\neq0$.

Assembly bias is one possible generative interpretation of this joint dependence.
Quantifying it uniquely requires additional assumptions regarding the definition of {$\theta$}, observable proxies, and the generative model.

Accordingly, the goal of this work is not to estimate assembly bias,
but to provide an implementation framework that separates estimation from interpretation, based on the observable quantities $g(r;m_1,m_2)$ and $\Delta_0(r;m_1,m_2)$.
While $\Delta_X(r;m_1,m_2)$ was introduced as a diagnostic for deviations from independence, it also serves as an indicator of the underlying joint structure.

\section{Numerical Illustration of Joint Mark Structure}
\label{sec:numerical_illustration}

\begin{figure}
    \centering
    \includegraphics[width=0.95\linewidth]{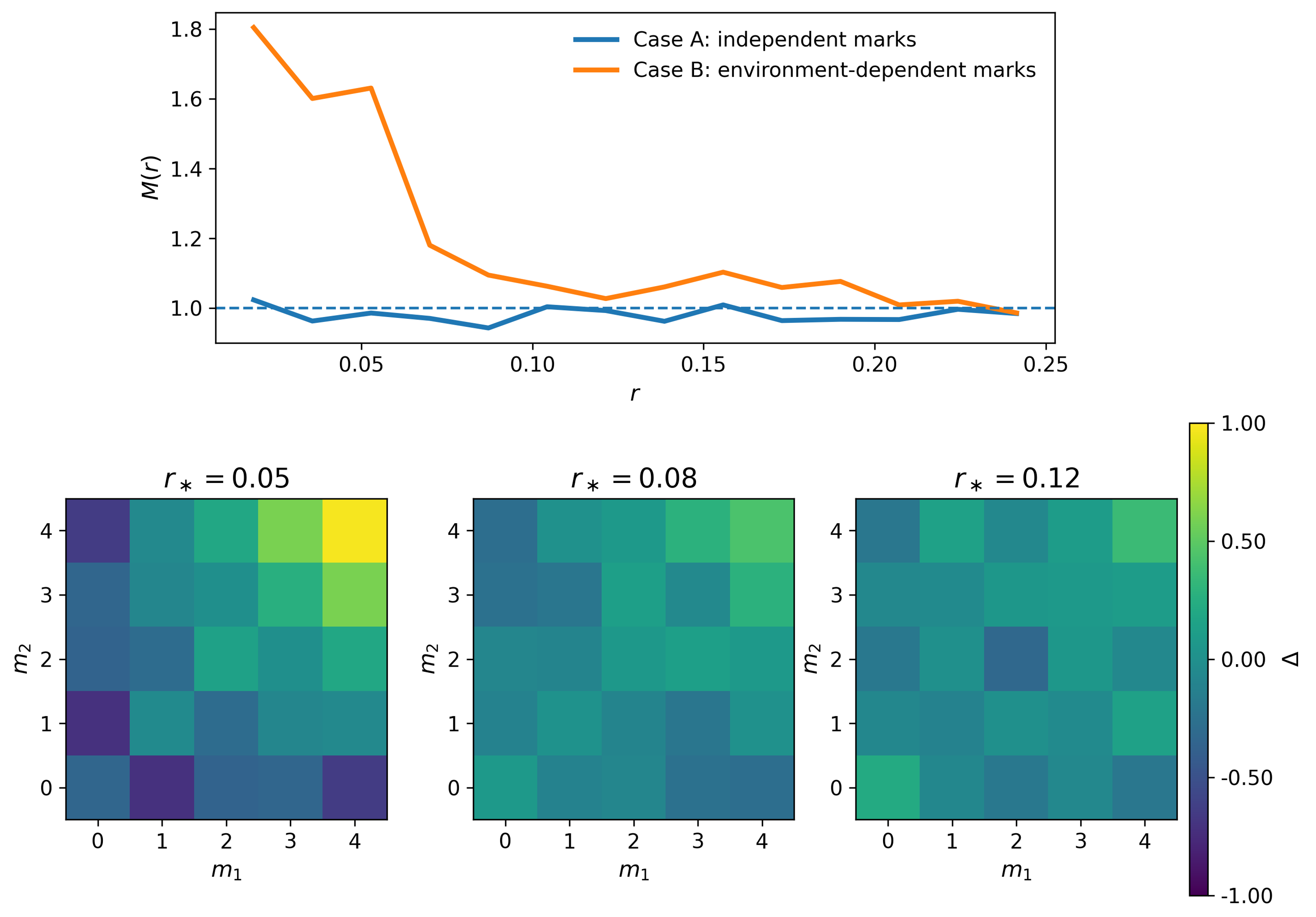}
    \caption{Comparison between independent mark assignment (Case~A) and environment-dependent mark assignment (Case~B) for the same spatial configuration.
    The top panels show the marked correlation $M(r)$, while the bottom panels display heatmaps of $\Delta_X(r_\ast;m_1,m_2)$ at multiple distance bins $r_\ast$.
    Although $M(r)$ tends to be enhanced at small scales in Case~B, $\Delta_X$ decomposes the contribution by mark pair and visualizes its scale dependence.
    }\label{fig:illustrating_example_multir}
\end{figure}

In this section, we demonstrate through a simple numerical experiment how the diagnostic quantity for joint structure introduced in this paper, $\Delta_X(r;m_1,m_2)$, visualizes mark dependence that is difficult to discern using projected statistics.

{
To clarify the setup, we briefly summarize the procedure used to generate the examples shown in this section. 
We consider a mock galaxy sample with prescribed spatial clustering and assigned marks. 
The spatial distribution is generated as a Thomas process (a Neyman--Scott cluster process) with parent intensity $\lambda_p=0.01$, mean number of offspring per parent $\mu=20$, and Gaussian dispersion $\sigma\sim 1$. 
Marks are then assigned to each point while keeping the spatial configuration fixed. 
In the null case (Case A), marks are drawn independently from the marginal distribution $p(m)$. 
In the environment-dependent case (Case B), the local density, defined as the number of neighbors within a sphere of radius $r_{\rm env}=5$, is used as an environmental indicator, and marks are assigned according to a prescribed functional form (linear or quadratic; see below). 
Note that all distances are expressed in arbitrary units.
}

{
For each realization, we estimate the joint pair correlation $g(r; m_1, m_2)$ and compute $\Delta_X(r; m_1, m_2)$ using binning in both separation and mark space. 
The resulting quantities are then visualized either as heatmaps on the $(m_1, m_2)$ plane or as functions of separation for representative mark pairs.
While the precise numerical choices affect quantitative details, the qualitative behavior discussed here is robust, and the procedure is fully reproducible given the definitions provided.
}

{
For comparison, we also use the conventional marked correlation
function $M(r)$ defined by eq.~\eqref{eq:mark_correlation_classical}.
}
Figure~\ref{fig:illustrating_example_multir} shows the marked correlation $M(r)$ for both cases, along with $\Delta_X(r_\ast;m_1,m_2)$ evaluated at multiple distance bins $r_\ast$.
While $M(r)$ exhibits an enhancement at small scales in Case~B, the heatmaps of $\Delta_X$ directly visualize which mark pairs contribute at which scales.
{
While $M(r)$ differs between the two cases, it provides only a
projected summary and does not identify the mark pairs responsible for the signal. 
In contrast, $\Delta_{\mathrm{ind}}(r;m_1,m_2)$ resolves this degeneracy by decomposing the joint structure in mark space.
}

\begin{figure}
    \centering
    \includegraphics[width=0.95\linewidth]{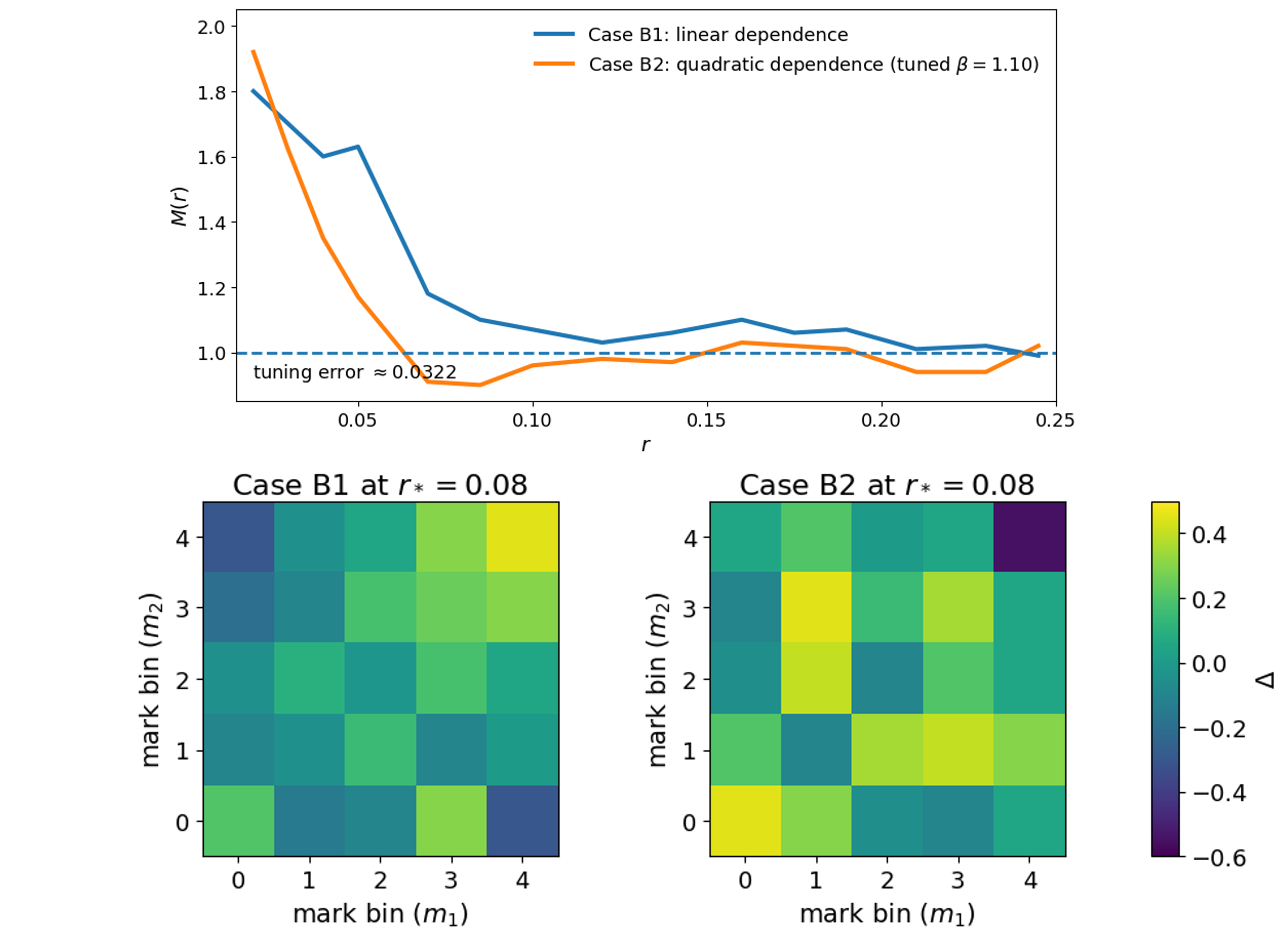}
    \caption{An example illustrating the limitation of projected statistics.
    We consider mark generation with linear dependence on environment (Case~B1) and quadratic dependence (Case~B2), adjusting the amplitude in the latter so that $M(r)$ closely matches.
    As shown in the top panels, {$M(r)$ alone does not reveal the origin of the difference in terms of specific mark pairs, whereas the bottom panels of $\Delta_X(r_\ast;m_1,m_2)$ clearly reveal distinct joint structures.}
    }\label{fig:illustrating_example_projection}
\end{figure}

Next, we construct an example in which different mark-generation mechanisms produce nearly identical $M(r)$.
In Case~B1, we assume a linear dependence on environment, while in Case~B2 we assume a quadratic dependence.
By adjusting the amplitude in the latter, $M(r)$ can be made to match that of Case~B1.
{
While $M(r)$ differs between the two cases, it provides only a
projected summary and does not identify the mark pairs responsible for the signal. 
In contrast, $\Delta_{\mathrm{ind}}(r;m_1,m_2)$ resolves this degeneracy by decomposing the joint structure in mark space.
As shown in Fig.~\ref{fig:illustrating_example_projection}, while the projected statistic $M(r)$ appears nearly indistinguishable at first glance, $\Delta_X(r_\ast;m_1,m_2)$ exhibits qualitatively different patterns.
}

These results show that while projected statistics may obscure differences in joint structure, $\Delta_X$ functions as a local diagnostic based on position-only clustering.

\section{Discussion}\label{sec:discussion}


From our standpoint, mark-dependent clustering is defined intrinsically as a joint structure that may involve continuous marks, and binned representations should be regarded as special cases thereof.
Below we organize the discussion in the following order:
(i) formulation with continuous marks,
(ii) basis expansion to control the estimation dimension,
and (iii) correspondence with binned representations.

{
The basis expansion introduced here provides a framework in which the analyst can explicitly control the degree of projection in mark space, thereby continuously bridging exploratory visualization via $\Delta_X$ and stable low-dimensional summaries.
An example of such statistics by \citet{CronieVanLieshout2016} is compared and discussed in Appendix~\ref{sec:discussion_J_correspondence}.
}

\subsection{Continuous-mark formulation and the dimensionality issue}
\label{subsubsec:continuous_mark}

In our formulation, mark-dependent clustering is most directly described by the joint pair correlation $g(r;m_1,m_2)$, or equivalently by the mark-pair distribution conditioned on distance $r$, $p(m_1,m_2\mid r)$.
Here the mark $m$ may be a continuous variable (e.g., absolute magnitude, stellar mass, SFR), and $(m_1,m_2)$ is a variable on a two-dimensional continuous space.
However, if one attempts to estimate this definition directly, one must infer a continuous bivariate distribution for each distance bin $r$, which is not realistic for finite-sample observational data.
This difficulty reflects the fact that marked clustering is intrinsically a high-dimensional problem involving the joint structure of distance $r$ and the mark pair $(m_1,m_2)$.

Conventional weighted marked statistics can be interpreted as avoiding this issue by implicitly projecting this high-dimensional structure.
In order to control this projection explicitly, we introduce a basis expansion in mark space.

\subsection{Controlling dimensionality: basis expansion in mark space}
\label{subsubsec:basis_expansion}

{
To bridge the gap between full distribution and stable summaries, we project the mark-pair space onto a set of basis functions \(\{\phi_a(m)\}_{a=0}^{A}\). 
The resulting statistics
\begin{align}
C_{ab}(r)
\equiv
\mathbb{E}\!\left[\phi_a(m_1)\phi_b(m_2)\mid r\right]
\label{eq:C_ab}
\end{align}
represent a controllable-dimensional summary of the joint correlation. 
Here, the expansion order $A$ functions as a tuning parameter: a higher $A$ preserves more physical information (e.g., fine features in the mass-SFR plane) at the cost of increased estimation variance.
}

As the order $A$ increases, more information along mark directions is retained, while estimation error also increases.
Thus $A$ acts as a tuning parameter that controls the trade-off between information content and stability.
From observational data, given the set of pairs in a distance bin $r$, $\mathcal{P}(r)$, one can estimate
{
\begin{align}
    \widehat{C}_{ab}(r) =
    \dfrac{1}{|\mathcal{P}(r)|}
    \sum_{(i,j)\in\mathcal{P}(r)} \phi_a(m_i)\phi_b(m_j)\label{eq:Cab_estimator}
\end{align}
so that distance binning and projection along mark directions are clearly separated.
These quantities can be viewed as linear projections of the conditional
mark-pair distribution onto the chosen basis.
}

\subsection{Complementary roles of $C_{ab}(r)$ and $\Delta_X(r;m_1,m_2)$}
\label{subsubsec:Cab_vs_Delta}

In this paper, we use both the projected statistics $C_{ab}(r)$ and the diagnostic $\Delta_X(r;m_1,m_2)$ to describe mark-dependent clustering.
Both are based on the joint correlation $g(r;m_1,m_2)$, but their roles differ.

$C_{ab}(r)$ is a summary statistic obtained by projecting the mark-pair distribution onto basis directions, compressing high-dimensional mark-dependent structure into a low-dimensional representation suitable for quantitative comparison.
In contrast,
\begin{align}
    \Delta_X(r;m_1,m_2) \equiv \ln\dfrac{g(r;m_1,m_2)}{g_X(r)}
\end{align}
is an exploratory diagnostic without projection, directly expressing which mark pairs are relatively over- or under-represented at each distance scale.

A natural workflow is therefore to first use $\Delta_X$ to identify scales and qualitative features where mark dependence exists, and then use $C_{ab}(r)$ to quantify the structure with controlled information content.

\subsection{Binned luminosity (or absolute-magnitude) representation as a special case of basis choice}
\label{subsubsec:continuous_vs_binned}

{The common practice of binning marks (e.g., luminosity samples) is naturally recovered as a special case of this basis expansion. 
By choosing bin indicator functions (eq.~\eqref{eq:bin_basis}) as the basis, $C_{ij}(r)$ simplifies to the conditional probability $\Pr(L_1\in A_i,\,L_2\in A_j\mid r)$. 
This identifies the traditional cross-correlation approach as a specific, discrete projection within our continuous-mark framework.
}

{
In simple terms, this compares luminosity-dependent clustering to the
overall clustering, while keeping track of which luminosity pairs contribute.
Consider restricting the mark to luminosity $L$ and partitioning it into finitely many bins $A_i$.
This corresponds to choosing the bin indicator functions
\begin{align}
\phi_i(L)
\equiv
\begin{cases}
1 & (L\in A_i),\\
0 & (\text{otherwise})
\end{cases}
\label{eq:bin_basis}
\end{align}
as basis functions.
Under this choice,
\begin{align}
    C_{ij}(r) = \mathbb{E}\!\left[\phi_i(L_1)\phi_j(L_2)\mid r\right]
    = \Pr(L_1\in A_i,\,L_2\in A_j\mid r) \label{eq:Cij_binned}
\end{align}
yielding a matrix of occurrence probabilities conditioned on distance $r$.
}

\subsection{Diagnosing magnitude dependence and the large-scale limit}

Writing the magnitude-bin correlation as
\begin{align}
\xi_{ij}(r)
\equiv
\xi\left(r;M_1\in A_i,\,M_2\in A_j\right),
\end{align}
we define
\begin{align}
\Delta_{X,ij}(r)
\equiv
\ln\dfrac{1+\xi_{ij}(r)}{1+\xi_X(r)}
\end{align}
which serves as a discrete version of $\Delta_X$ localized to bin pair $(i,j)$ and directly visualizes departures from the separation hypothesis.

At sufficiently large scales,
\begin{align}
\xi_{ij}(r)
\simeq
b_i\,b_j\,\xi_m(r)
\end{align}
holds, and the magnitude-bin correlation matrix has an approximately rank-1 structure.
This indicates that detailed joint structure at small scales reduces to an effective bias description at large scales, illustrating that the present framework enables a scale-spanning understanding.

\section{Conclusions and Outlook}\label{sec:summary_outlook}

In this paper, we formulated marked galaxy clustering by treating galaxies as points on the product space $\mathbb{R}^3\times\mathcal{M}$ and by taking the joint pair correlation $g(r;m_1,m_2)$ (or equivalently the separation-conditioned mark-pair distribution $p(m_1,m_2\mid r)$) as the fundamental quantity.
From this perspective, commonly used marked correlation functions can be understood in a unified way as projections of an underlying joint structure.
The key contribution of this work is to provide an explicit framework in which marks and clustering are treated jointly at the definitional level, and to show that reweighting operations do not alter the joint correlation itself but only its projection.
We further demonstrated that violations of the separation hypothesis can be directly diagnosed through $\Delta_X(r;m_1,m_2)$.
These results follow naturally from the basic quantities of point process theory and do not rely on particular weight choices or model assumptions.

Within this framework, phenomena traditionally discussed as independent effects—such as marked clustering signals and assembly bias—are more naturally interpreted as projections of an underlying joint dependence structure.
Because observational statistics are typically marginals of latent generative processes, distinct dependence structures may yield identical projected observables.
In this sense, assembly bias cannot be uniquely inferred without additional model assumptions.
{What can be directly diagnosed from observations is the presence of non-factorizable joint structure,
\begin{align}
    g(r;m_1,m_2)\neq g_X(r)\,p(m_1)p(m_2),
\end{align}
or equivalently $\Delta_{\mathrm{ind}}(r;m_1,m_2)\neq0$.}

Using luminosity-dependent clustering as an illustrative example, we showed that the classical statement that brighter galaxies cluster more strongly can be expressed directly in terms of the $M$-dependence of the joint correlation, or equivalently through the structure of magnitude-binned correlation matrices.
This viewpoint makes explicit which magnitude pairs contribute at which scales, information that is difficult to recover from a single weighted marked statistic.

Although the present work focuses on the definition and interpretation of observational statistics, the framework also admits a natural generative interpretation.
If galaxy marks are assigned according to a local environment $\Phi$ (e.g., density field, tidal field, cosmic web classification, or halo mass proxy), a mark-assignment kernel $p(m\mid x;\Phi)$ may be introduced, and the joint correlation $g((x_1,m_1),(x_2,m_2))$ can then be viewed as an observational constraint on such models.
Projected statistics take the unified form
\begin{align}
    S_w(r)=\dfrac{\displaystyle \int w(m_1,m_2)\,g(r;m_1,m_2)\,\pd m_1\pd m_2}{\displaystyle \int g(r;m_1,m_2)\,\pd m_1\pd m_2},
\end{align}
allowing the design of weight functions to be separated from the estimation of the joint structure.
While connections to forward modeling and Bayesian inference lie beyond the scope of this work, adopting the joint correlation as the fundamental quantity provides a natural starting point for such developments.

Finally, the formulation can be extended straightforwardly from luminosity to stellar mass, star formation rate, morphology, or even multivariate property vectors.
By controlling the effective dimension through basis expansion or binning while retaining the joint structure between marks and environment, the approach offers a flexible and information-preserving framework for analyzing future large-scale survey data.

\section*{Acknowledgments}
We deeply thank the referee, Prof.\ Zheng Zheng, for careful reading of the manuscript and constructive suggestions that improved the clarity of this paper very much.
We are grateful to Shiro Ikeda, Satoshi Kuriki, Keisuke Yano, Atsushi J.\ Nishizawa, and Shun-ya S.\ Uchida for valuable discussions and insightful comments related to this work.
This research was supported by JSPS Grant-in-Aid for scientific research (24H00247), and by the joint research program of the Institute of Statistical Mathematics (General Research 2), ``Machine-Learning Cosmogony: From Structure Formation to Galaxy Evolution.''

\section*{Data Availability}
 
The models and analysis code in this paper will be shared upon reasonable request to the authors.



\bibliographystyle{mnras}
\bibliography{mark_correlation} 



\appendix

\section{Connection Between Reweighting and Palm Measures}\label{app:palm}

{
The appendix provides additional theoretical context for the framework developed in the main text. 
The first part clarifies the relation to standard concepts in point process theory, while the second part illustrates how summary statistics can be interpreted as projections of the joint structure. 
}

Here, we clarify how the reweighting operation introduced in this work corresponds to Palm measures, size-biased point processes, and tilting (change of measure) in point process theory.
{
We denote by $\nu$ a reference measure on the mark space (e.g.\ the marginal mark distribution), and by $B(0,r)$ the ball of radius $r$ centered at the origin in $\mathbb{R}^3$. 
These are standard notations in spatial statistics and point process theory.}
For details, see
\citet{Ripley1981SpatialStatistics},
\citet{2003Daley_point_processI,2008Daley_point_processII},
\citet{moller2003statistical},
and \citet{baddeley2015spatial}.

\subsection{Palm Measures and Conditional Distributions}

For a general marked point process, the distribution of other points conditional on the presence of a point with a given location and mark is defined by the Palm measure.
The mark-pair distribution conditioned on separation, $p(m_1,m_2\mid r)$, can be interpreted as the conditional distribution associated with the two-point Palm measure.
Accordingly, the quantities used in the main text,
\begin{align}
    \rho^{(2)}(r;m_1,m_2), \qquad p(m_1,m_2\mid r),
\end{align}
are consistent with the standard objects of Palm theory.

\subsection{Reweighting and Size Bias}

Given a nonnegative weight function $u(m)$, the reweighted point process defined by
\begin{align}
    \rho_u^{(1)}(x,m)=u(m)\rho^{(1)}(x,m)
\end{align}
corresponds to introducing size bias with respect to the marks.

This operation can be interpreted as a natural generalization of size-biased sampling, in the sense that the probability of observing a point with mark $m$ is modified proportionally to $u(m)$ \citep[e.g.][]{moller2003statistical,baddeley2015spatial}.

Crucially, this reweighting does not alter the spatial configuration of points itself, but only changes which points (or pairs) are statistically emphasized.
Therefore, the joint pair correlation $g(r;m_1,m_2)$ remains invariant under such reweighting.

\subsection{Interpretation as Tilting (Change of Measure)}

From a measure-theoretic perspective, reweighting can be understood as tilting (change of measure) of the original point process distribution.
That is, it corresponds to transforming the probability measure as
\begin{align}
    \pd \mathbb{P}_u \propto \prod_i u(m_i)\,\pd \mathbb{P}.
\end{align}
Under this transformation, the joint structure conditioned on separation remains invariant as $g(r;m_1,m_2)$, and the positional clustering $g_X(r)$ is likewise unchanged\footnote{
Throughout this paper, reweighting is used as an \emph{estimator-level} operation (importance weighting of points/pairs) applied to a fixed observed catalog, rather than as a procedure that generates a new ``tilted'' location process.
In this sense, the invariance statements concern the \emph{normalized joint} pair correlation on $\mathbb{R}^3\times\mathcal{M}$ (i.e.\ ratios in which the factors $u(m_1)u(m_2)$ cancel), whereas the \emph{marginal} location intensity/correlation obtained after integrating out marks may change under an actual change of measure unless additional assumptions are imposed.}.

Consequently, the diagnostic quantity introduced in the main text,
\begin{align}
    \Delta_X(r;m_1,m_2) \equiv \ln\dfrac{g(r;m_1,m_2)}{g_X(r)},
\end{align}
is also invariant under the tilted measure.
From this viewpoint, conventional marked correlation functions amount to observing projections of low-order moments under a tilted measure.

In summary, the reweighting framework introduced in this work is fully consistent with standard concepts in point process theory such as Palm measures, size bias, and tilting, and provides theoretical support for the invariance of the joint correlation $g(r;m_1,m_2)$ and the diagnostic quantity $\Delta_X(r;m_1,m_2)$.

{\section{Relation to Summary Statistics in Spatial Statistics}
\label{sec:discussion_J_correspondence}
}

{
In this Appendix, we briefly discuss the relation between the present
framework and summary statistics developed in spatial statistics,
in particular those introduced by \citet{CronieVanLieshout2016}.
}
They defined an inhomogeneous cross$J$-function for marked point processes under the assumption of intensity-reweighted moment stationarity (IRMS)\footnote{
Intensity-reweighted moment stationarity (IRMS) is an assumption requiring that, even when the intensity of a point process is spatially inhomogeneous, the $n$-point correlation functions and factorial moment measures become translation invariant after appropriate intensity reweighting.
Under this condition, one can remove position-dependent intensity fluctuations and define and compare the correlation structure itself, so that summary statistics such as the $K$-function or the cross$J$-function remain meaningful in inhomogeneous settings.
In contrast, the focus of this paper is that we do not take such a reweighting assumption as a premise, but instead adopt the joint pair correlation on the direct product space as the fundamental quantity.
}.
For mark sets $C,E\subset M$, they expressed it in terms of $n$-point correlation functions $\{\xi_n\}$ as
\begin{align}
  J_{\mathrm{inhom}}^{C,E}(r)
  =
  \dfrac{1}{\nu(C)}
  \left[
    \nu(C)
    +\sum_{n=1}^{\infty}\dfrac{(-\bar\lambda_E)^n}{n!}\,J_n^{C,E}(r)
  \right],
  \qquad (r\ge 0)
  \label{eq:J_inhom_series}
\end{align}
\citep[][Definition~2, eq.~(11)]{CronieVanLieshout2016}.
Following their notation,
\begin{align}
  &J_n^{C,E}(r) \notag \\
  &\quad =
  \int_{b\in C}
  \int_{(B(0,r)\times E)^n}
  \xi_{n+1}\left((a,b),(z_1+a,m_1),\ldots,(z_n+a,m_n)\right)\,\notag \\
  &\qquad \times  \prod_{i=1}^{n}\{\pd z_i\,\pd \nu(m_i)\}\,
  \pd \nu(b)
\end{align}
where $\bar\lambda_E$ is a constant defined from the minimum intensity on the $E$ component.
In \citet{CronieVanLieshout2016}, $\xi_n$ are defined recursively as intensity-reweighted densities of factorial cumulants (with $\xi_1\equiv 1$), and for a Poisson process one has $\xi_n\equiv 0$ for $n\ge 2$.

\paragraph{(i) Identifying our $\xi(r;m_1,m_2)$ with $\xi_2$ in \citet{CronieVanLieshout2016}}

We adopt $\xi=g-1$, whereas Cronie \& van Lieshout introduce $\xi_2$ as the second-order product density $\rho^{(2)}$ normalized by intensities (as follows from the definition of $\xi_n$ and its relation to $\rho^{(n)}$; see also \citealt{CronieVanLieshout2016}).
Thus, for the two-point case, a simple notational correspondence yields
\begin{align}
  g(r;m_1,m_2) &=1+\xi_2(r;m_1,m_2) \notag \\
  \qquad&\Longleftrightarrow\qquad
  \xi(r;m_1,m_2)=\xi_2(r;m_1,m_2).
\end{align}
In other words, from our perspective, the $J$-function of Cronie \& van Lieshout is a \emph{summary operator acting on the joint correlation $\xi(r;m_1,m_2)$ (and higher-order $\xi_{n\ge 3}$)}.

\paragraph{(ii) Reduction map from $\Delta_X$ to $J$ (removal of mark locality)}

$J_{\mathrm{inhom}}^{C,E}(r)$ does not retain the specific mark values $(m_1,m_2)$; instead it aggregates joint information by averaging over the sets $C$ and $E$.
In our notation, define the mark-set-averaged joint correlation as
\begin{align}
  \bar \xi^{C,E}(r)
  \equiv
  \dfrac{1}{\nu(C)\nu(E)}
  \int_{C}\int_{E}
  \xi(r;m_C,m_E)\,
  \pd \nu(m_C)\,\pd \nu(m_E)
  \label{eq:xi_bar_CE}
\end{align}
(where $C,E$ may be interpreted as ``magnitude bins,'' etc.).
Using $\xi=g-1$ and $\Delta_X=\ln(g/g_X)$, if $g_X(r)$ is independent of marks, then
\begin{align}
  \bar \xi^{C,E}(r) &=
  \dfrac{1}{\nu(C)\nu(E)}
  \int_{C}\int_{E}
  \left[g_X(r)\exp\{\Delta_X(r;m_C,m_E)\}-1\right]\, \notag \\
  &\quad \times \pd \nu(m_C)\,\pd \nu(m_E)
  \label{eq:xi_bar_from_Delta}
\end{align}
showing that the local structure of $\Delta_X$ in mark space is reduced to an average over $C$ and $E$.
Already at this stage, the locality information carried by $\Delta_X(r;m_1,m_2)$, namely, \emph{which mark pairs contribute}, is lost.

\paragraph{(iii) Reduction map from $\Delta_X$ to $J$ (geometric eventization and mixing of higher-order hierarchy)}

Moreover, $J_{\mathrm{inhom}}^{C,E}(r)$ is constructed (a) from geometric events such as nearest-neighbor distances and voids, and (b) consequently as an all-orders series involving not only the two-point correlation but also $\xi_{n\ge 3}$ (eq.~\eqref{eq:J_inhom_series}) \citep[][]{CronieVanLieshout2016}.
In this sense, one may interpret that a multi-stage reduction (information compression) is performed:
\begin{align}
  \Delta_X(r;m_1,m_2)
  &\;\;\xrightarrow[\text{set averaging}]{(m_1,m_2)\mapsto (C,E)}\;\;
  \bar\xi^{C,E}(r) \notag \\
  &\;\;\xrightarrow[\text{geometric eventization + all-order summary}]{}\;\;
  J_{\mathrm{inhom}}^{C,E}(r).
\end{align}

\paragraph{(iv) Correspondence at low order (two-point approximation): first-order expansion of $J$}

Truncating the series (eq.~\eqref{eq:J_inhom_series}) at $n=1$ yields
\begin{align}
  J_{\mathrm{inhom}}^{C,E}(r)
  &\approx 1 - \dfrac{\bar\lambda_E}{\nu(C)}\,J_1^{C,E}(r) \notag \\
  &\qquad (r \ \text{small, or in the weak-correlation limit})
  \label{eq:J_first_order}
\end{align}
\citep[][eq.~(11)]{CronieVanLieshout2016}.
Since $J_1^{C,E}$ is an integral of $\xi_2$ over $B(0,r)\times E$, in our notation it can be understood conceptually as
\begin{align}
  J_1^{C,E}(r) &\sim  \nu(C)\,\nu(E)\,
  \int_{B(0,r)}\bar\xi^{C,E}(|z|)\,\pd ^3 z \notag\\
  &=
  4\pi\,\nu(C)\,\nu(E)\,\int_{0}^{r}\bar\xi^{C,E}(t)\,t^2\,\pd t
  \label{eq:J1_xibar_relation}
\end{align}
(where we assumed isotropy so that the integrand depends only on $|z|=t$).
Thus, the low-order approximation of $J$ corresponds to a spherical average of the set-averaged two-point correlation from $C$ to $E$, and is determined via $\Delta_X$ through eq.~\eqref{eq:xi_bar_from_Delta}.

\paragraph{(v) Why $J$ alone is insufficient for the astronomical goals}

The inhomogeneous cross$J$-function $J_{\mathrm{inhom}}^{C,E}(r)$ of \citet{CronieVanLieshout2016} is useful as a summary statistic that stably detects deviations from a Poisson (uncorrelated) baseline.
However, its purpose differs from the main goals of this paper: \emph{reconstructing the joint structure of mark pairs conditioned on distance $r$} and \emph{directly describing which mark combinations are over- or under-represented at which scales}.
As emphasized above, the design philosophy of summary statistics can obscure how much joint information is retained, and makes it difficult to express relations of interest (e.g., luminosity-dependent bias or morphology-dependent clustering) at the level of fundamental quantities.
Our diagnostic $\Delta_X(r;m_1,m_2)$ is introduced precisely to fill this gap:
while $J$ is powerful for ``detection'' through multi-stage aggregation,
$\Delta_X$ provides a non-projected description that retains the local structure of contributing mark pairs.

\medskip
{
In summary, the constructions in this appendix provide a theoretical interpretation of the framework developed in the main text. 
The joint pair correlation $g(r;m_1,m_2)$ and the diagnostic $\Delta_X(r;m_1,m_2)$ can be understood as fundamental quantities on the product space, while conventional summary statistics correspond to various forms of projection or aggregation of this structure.
This perspective clarifies how the present approach complements existing methods by retaining information that is typically lost in
projected statistics.
}

\bsp	
\label{lastpage}
\end{document}